\documentclass[10pt,a4paper]{article} 
\pdfoutput=1 

\usepackage{amsmath}

\usepackage{caption}

\usepackage[T1]{fontenc}
\usepackage{graphicx}
\usepackage{tikz}

\usepackage{subcaption}

\usepackage[margin=2.5cm]{geometry}

\usepackage[pdftex, colorlinks=true, linkcolor=blue, urlcolor=blue]{hyperref}

\begin{document}
\title{Spatial Process Mining}

\author{
    Shintaro Yoshizawa\textsuperscript{1} \and
    Takayuki Kanai\textsuperscript{1} \and
    Masahiro Kagi\textsuperscript{1}
}\date{} 

%

\footnotetext[1]{R-Frontier Division, Toyota Motor Corporation, Toyota, Japan.
Email: \{first\_lastname\}@mail.toyota.co.jp}

\maketitle              
\begin{abstract}
We propose a new framework that focuses on on-site entities in the digital twin, a pairing of the real world and digital space. Characteristics include active sensing to generate event logs, spatial and temporal partitioning of complex processes, and visualization and analysis of processes that can be scaled in space and time. As a specific example, a cell production system is composed of connected manufacturing spaces called cells in a manufacturing process. A cell is sensed by ceiling cameras to generate a Gantt chart that provides a bird's-eye view of the process according to the cycle of events that occur in the cell. This Gantt chart is easy to understand for experienced operators, but we also propose a method for finding the focus of causes of deviations from the usual process without special experience or knowledge. This method captures the characteristics of the processes occurring in a cell by using our own event node ranking algorithm, a modification of HITS (Hypertext Induced Topic Selection), which scores web pages against a complex network generated from a process model. \\


\end{abstract}
\section{Introduction}
Currently, process mining techniques are widely used in all industries to visualize and improve processes, but in the real world, one of the main challenges is the preparation and maintenance of data, taking into account collection and maintenance costs. Moreover, even if data is available to generate event logs, it is not easy to perform analysis of processes involving multiple processes. In response to this problem, research on object-centric process mining is currently being promoted~\cite{ref_wil1}.

Object-centric process mining is a method of process mining in an abstract cyberspace that utilizes data stored in databases, etc. However, in order to grasp complex processes more easily, we are considering process mining from the point of actively sensing events that exist in physical spaces such as actual homes, workplaces, public facilities, distribution sites, and cities and so on. To do so, we are focusing on the size and resolution of space and time in which entities such as people and objects exist. If necessary, the robot can be equipped with sensors. In other words, we use physical space as the starting point for mining. We call the approach of dividing the physical space, integrating the event logs of entities in the divided space, and mining them "{\it Spatial Process Mining}".

\begin{figure}[htbp]
\centering
\includegraphics[width=0.75\textwidth]{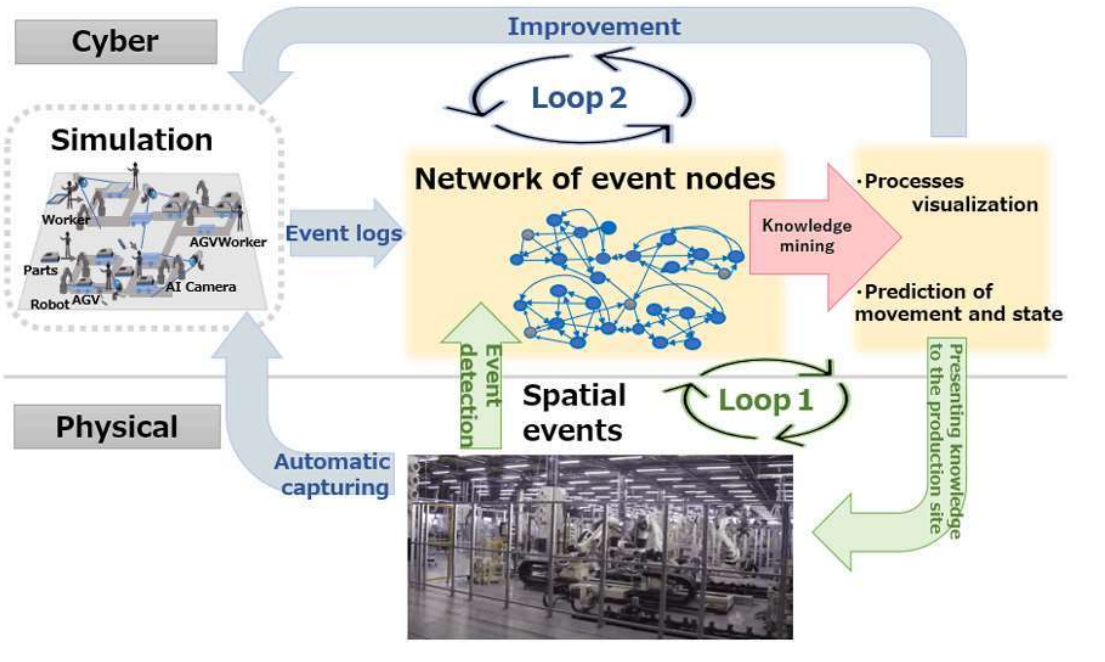}
\caption{Digital Twin with Double Loops.} 
\label{Fig1}
\end{figure}

Figure~\ref{Fig1} is an example of a digital twin in a manufacturing environment. 
Spatial process mining has a first loop that feeds back the results and findings to entities in physical space, and a second loop that feeds back the analyzed results to process design and optimization studies. In Loop 1, the highly skilled person expects to be able to easily understand the whole situation simply by being able to overlook the facts and the ability to respond to the unexpected. In addition, the first loop captures the time required for each entity event, which is used as input information for design and optimization studies, and the second loop, realistic planning and simulation-based forecasting of effects, can be performed. Since the number of pages is limited, this paper will focus on Loop 1.

\begin{figure}
\centering
\includegraphics[width=0.75\textwidth, keepaspectratio]{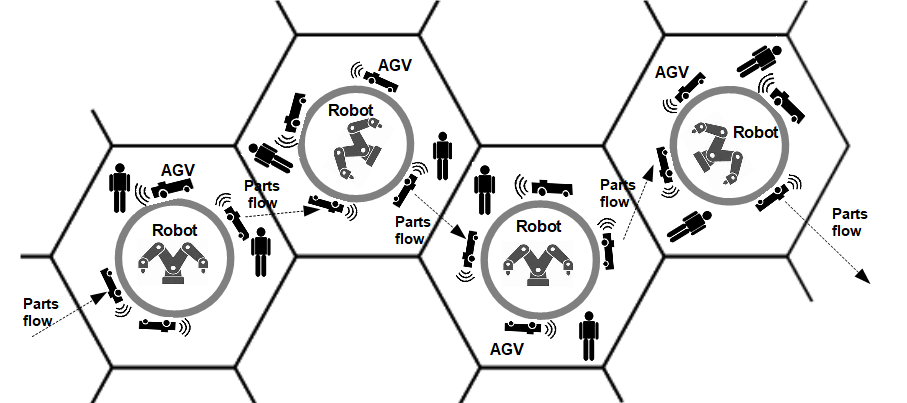}
\caption{Example of a conceptual picture of a cell production system combining hexagonal cells.} \label{Fig2}
\end{figure}

Considering, for example, the home as a small space and the city as a large space as the physical space of the digital twin, we chose {\it \textbf{the cell production system}} 
 (Fig.~\ref{Fig2}), the site of a production process that falls somewhere in between in terms of scale.
A cell production system is modularly composed by combining production activities among cells. The transfer of parts between cells is performed by workers using equipment such as loading assistance arms or special dollies. We thought that a cell production system would be suitable for studying this spatial scalability. Entities such as multiple workers and multiple automated guided vehicles (AGVs) that appear in production activities within a cell are characterized by their activities being synchronized and asynchronized with a certain degree of flexibility. The larger the target space, the more complex the process becomes, and the more difficult it becomes to evaluate the entire process. Therefore, we grasp the "usual process", consider changes from the standard based on this standard, and perform "KAIZEN(Improvement)" by on cite workers in the physical space and by planner in the cyber space.

An example of active sensing is the increasing use of computer vision~\cite{ref_Luft,ref_vision}. Computer vision is an excellent low-cost mechanism for actively collecting events that occur in the time and space one wishes to observe. Object detection models are now mature and readily available for general use~\cite{ref_yolov3,ref_opencv}. However, object detection techniques have accuracy problems depending on the sensing conditions. The challenge is to train detection models based on knowledge of the characteristics of object detection models. In cell production, a specific case of detecting entities as objects in a certain cell with two cameras and generating event logs from the detection results is described in the text.

One of the typical measures of a production process is {\it \textbf{the production cycle time}}. In cell production, the "usual cycle time" can be estimated for this measure, but when the cycle time becomes longer, we would like to know the role of event nodes and the factors that have caused the cycle time to become longer

Assuming that a process model such as a directly follows graph~\cite{ref_dfg,ref_dfm} is obtained, we propose a simple method for discovering the role (function and importance) of event nodes by means of an algorithm that solves the eigenvalue problem of a matrix created from the process model. The background of our method is as follows.  Petri net was invented by Carl Adam Petri in the 1960s, and DSM (Design Structure Matrix) was proposed by Donald Steward at the same time (See for example~\cite{ref_dsm}). The relationship between Petri net and DSM was studied by Karniel and Reich~\cite{ref_pdsm}. In a different context from these studies, the Analytic Hierarchy Process (AHP) was proposed by Thomas L. Saaty in the 1970s as a decision-making method based on mathematics and psychology. 

The AHP has been extended to ANP (Analytic Network Process)~\cite{ref_anp}. DSM is a visual representation of the dependencies among the elements in a system, while ANP is a method of making decisions while considering these dependencies, although there are structural similarities between the two, the scope and purpose of their application differ. Moreover, Web search algorithms on the WWW(World Wide Web) were developed in the late 1990s, with PageRank~\cite{ref_pr}, and HITS (Hypertext Induced Topic Search) ~\cite{ref_hits} and so on (See, for example, ~\cite{ref_prbook,ref_no1}) .

DSM, AHP, ANP, and HITS have similar structures as problems to find the standard form of a matrix. More specifically, AHP, ANP, HITS, and PageRank are based on the power method, and the theoretical basis for these algorithms is supported by the Perron-Frobenius theorem. In these algorithms, the efficient ranking of the importance of network nodes and the change in ranking when nodes are added, especially the inversion of ranking, are discussed~\cite{ref_prbook,ref_no1}. These algorithms need to consider the hyperparameter related to  "Google's teleportation"  in order to avoid problems caused by the network topology (in general, any two nodes are not connected by a directed edge), but our proposed method does not need to consider the hyperparameter, and its theoretical basis is guaranteed by the theory of gradient dynamical systems.

We do not simply focus on the rankings, but rather on discovering "differences from the usual" by the characteristics of the distribution of the event node rankings. We have confirmed through the case of cell production that the variation of ranking distribution expresses "the degree of anxiety of the process". This concept would be broadly useful for complex process analysis. Details are given in the main text.

Section 2 below describes the elements of spatial process mining, Section 3 describes a camera-based event log generation method, Section 4 compares the node ranking method for conventional methods, Section 5 describes an application of spatial process mining to the cell production system, and Section 6 discusses the potential of spatial process mining in light of its application to cell production, Section 7 summarizes and comments future issues.

\section{Components of Spatial Process Mining}
The means to realize the double loop of the digital twin is spatial process mining, and Figure 3 shows the elements that make up the means. The double loop means that there are two major types of feedback (KAIZEN actions) to the final site. In more detail, as shown in Figure 3, there are three feedback destinations for the Digital Twin: process design, process modeling, and on-site entities.
\begin{itemize}
\item
 Process design consists of layout design and process procedure planning.
\item 
In process modeling, a Directly Follows Graph is automatically generated from event log data, and the process procedures are reviewed after determining the layout changes of work locations and the time required for elemental operations.
\item 
The target on-site entities (people, moving vehicles, robots, etc.) have two major feedback loops: a process optimization loop and a sensing and analysis loop.
\end{itemize}

This paper has two major original contributions. The first is the part of event detection and visualization based on sensing the detection target by area and spatial and temporal resolution. The second is the part related to the ranking analysis of event nodes in the analysis and state estimation (deviation from the usual state) utilizing the analysis results.

The next section describes the concept of the event log generation method using cameras in particular.

\begin{figure}
\includegraphics[width=\textwidth]{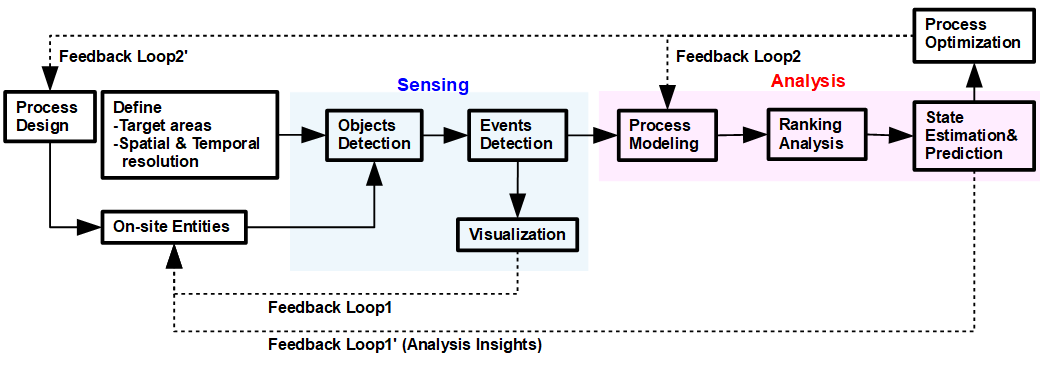}
\caption{Spacial Process Mining Components and Feedback Loops.} \label{Fig3}
\end{figure}

\section{Event Log Generation by Multiple Cameras}
This section describes the concept of event detection using multiple cameras and how to generate event logs. First, we discuss multi-object detection by multiple cameras as a pre-processing step for event detection, and then describe two methods for event detection.

\subsection{Multi-object detection by multiple cameras}
Since multiple cameras are deployed, there is overlap in the imaging area, and as a result, different cameras may image the same object. There are two approaches in this situation: \\
\textbf{Integration 1}: After merging the images from each camera into a single image (Image Stitching~\cite{ref_is}), object detection~\cite{ref_mot} is performed, and followed by event detection. \\
\textbf{Integration 2}: Object detection~\cite{ref_mot} at a specific location is performed on each camera image, and followed by event detection. The event detection results created for each camera are then merged, taking into account overlaps.

The main advantage of Integration 1 is the ease of annotation of the training dataset. Also, image-integrated videos are easier to comprehend. The main disadvantage is that stitching becomes more challenging depending on the camera position and orientation, and camera calibration is required.
The main advantage of Integration 2 is that the cameras share the name of the location to be detected, so camera calibration is not required. Furthermore, it is easy to add additional cameras to the system. The disadvantage of Integration 2 is that it requires time and effort to verify the detection accuracy since the detection target appears as multiple images. In this paper, we took the Integration 2 approach as the first trial.

\subsection{Event log format}
In spatial process mining, for a target area, the process analyst considers at which locations the event logs to be acquired can detect events and specifies locations. The specified location is considered to have location-specific features, such as specific tasks. Let $si$ denote the specified location and $\{si\}_{i =1, ..., N}$ denote the set of specified locations. Multiple cameras are positioned so that entities (people, moving objects, etc.) can be detected as bounding boxes on at least a specified set of locations. Entities are denoted by $Ej$, and $\{Ej\}_{j = 1,.... , M}$ denote the set of entities. The specific event log format is defined as follows:
\\
\{$si_1$, ($Ei_{1,1}$, property of $Ei_{1,1}$),... , ($Ei_{1,p}$, property of $Ei_{1,p}$);
$\cdots$ ; \\
\phantom{x} $si_n$, ($Ei_{
n,1}$, property of $Ei_{n,1}$),... , ($Ei_{n,q}$, property of $Ei_{n,q}$), timestamp \}.   
\\
Note that if events occur simultaneously, multiple locations $si_1,...,si_n$ are logged in an event log. 
Where $\{i_1,...,i_n\} \subseteq \{1,..., N\}$.

\begin{figure}
\includegraphics[width=\textwidth]{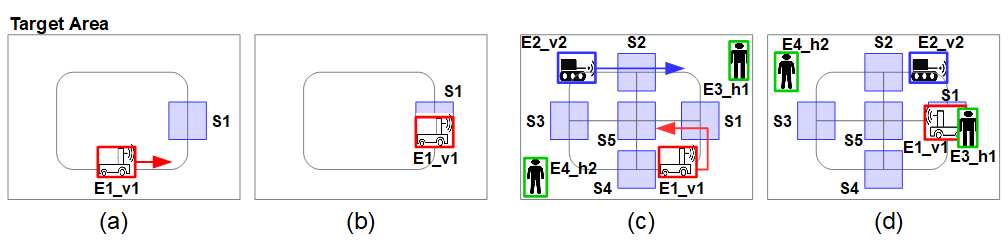}
\caption{Examples of Event Log Design.} \label{Fig4}
\end{figure}

For example, (a) in Figure 4 assume that entity $E1$ moves to location $s1$ as vehicle $v1$ and stops as shown in (b) in Figure 4 when it travels counterclockwise. At this time, when the red bounding box of the vehicle $v1$ overlaps with the blue area of $s1$ “under certain conditions, or when there is no overlap,” an event node at $s1$ is considered to have occurred. The vehicle $v1$ is in its first round (one cycle) and the event log is,
EL1: \{$s1, (E1,v1)$, 2024/08/15/17:40:50\}, etc. Where 2024/08/15/17:40:50 means Year/Month/Day/Hour/Minute/Second. The event log for the second round (second cycle) is denoted as EL2.
Entity $E1$ is $v1$, so it may be abbreviated as EL1: \{$v1\_s1$, 2024/08/15/17:40:50 \}. 

Figure 4 (c) shows a situation where the event log is not generated, but rather four entities consisting of two automated guide vehicles (AGVs) $E1\_v1$ and $E2\_v2$, two workers $E3\_h2$ and $E4\_h2$, and the AGVs are scheduled to move along an arrow, respectively. In Figure 4 (c), at location $s1$, the two entities overlap, and the event log for the first measurement is given as,\\
EL1:\{$s1,(E1,v1), (E3,h1)$, 2024/08/15/18:12:20\}. On the AGV, $E1\_v1$ stopped at location $s1$,
if it is clear that the worker $E3\_h1$ is to perform a given task, it may be abbreviated as EL1: \{$h1\_s1$, 2024/08/15/18:12:20 \}.

\subsection{Spatio-Temporal Resolution}
The period of logging should be specified taking into account the spatial arrangement between the specified locations $\{si\}$ (the spacing is not too narrow compared to the size of the bounding box for detecting people) and the speed and amount of movement of the entities (whether or not small instantaneous movements need to be detected). These settings depend on the object detection accuracy of the image recognition technology.

As described above, by specifying the locations where events are observed, event logs can be used to detect events occurring in various physical spaces, depending on the interference conditions between entities and the specified locations.

\section{Ranking Algorithms}
As mentioned in the introduction, basic ranking algorithms include $\textbf{HITS}$ and $\textbf{PageRank}$, but these methods use the maximum eigenvalue $\lambda_{max}$ of \textbf{the authority matrix}, $A = L^TL$ or \textbf{hub matrix}, $H = LL^T$ for the link matrix $L$ defined from the weighted network and its eigenvector $x=(x_1,.... ,x_n)^T$ to obtain the ranking of nodes $x_i,(i=1,...,n)$. That is $Ax_A=\lambda_{A, max} x_A$ and $Hx_H=\lambda_{H, max} x_H$.

The ranking differs depending on the structure of the link matrix, matrix element values, and the method used to solve the eigenvalue problem. When the size of the link matrix is large or the link matrix has dense non-zero elements, the conventional method requires consideration of hyperparameter adjustment. To address this problem, we propose a new node ranking method that can be handled simply.

Figure~\ref{Fig5} and Figure~\ref{Fig6} shows concrete examples of weighted networks. The link matrix $L_{[0]}$ in Figure~\ref{Fig5} and the link matrix $L_{[1]}$ in Figure~\ref{Fig6} are as follows.

\[ L_{[0]} =  
\begin{pmatrix}
  1.01 & 0.01 & 0.00 \\
  0.01 & 1.00 & 0.00 \\
  0.00 & 0.00 & 0.90 
\end{pmatrix}, 
\qquad
L_{[1]} = 
\begin{pmatrix}
  1.01 & 0.01 & 0.00 & 0.01 \\
  0.01 & 1.00 & 0.00 & 0.02 \\
  0.00 & 0.00 & 0.90 & 1.00 \\
  0.01 & 0.01 & 0.02 & 0.05 
\end{pmatrix}
\]

\begin{figure}
    \centering
    \begin{minipage}{0.3\textwidth}
        \centering
        \begin{tikzpicture}
           
            \node[circle, draw] (x2) at (0, 0) {$x_2$};
            \node[circle, draw] (x1) at (2, 2) {$x_1$};
            \node[circle, draw] (x3) at (2, -2) {$x_3$};

            \draw[->] (x1) to[bend left] node[midway, above] {0.01} (x2);
            \draw[->] (x2) to[bend left] node[midway, below] {0.01} (x1);
            \draw[->] (x1) edge[loop above] node {1.01} (x1);
            \draw[->] (x2) edge[loop above] node {1.00} (x2);
            \draw[->] (x3) edge[loop above] node {0.90} (x3);
        \end{tikzpicture}
        \caption{Weighted network of 3 nodes.}
         \label{Fig5}
    \end{minipage}%
    \hspace{2.0cm} 
    \begin{minipage}{0.3\textwidth}
        \centering
\begin{tikzpicture}
    \node[circle, draw] (x2) at (0, 0) {$x_2$};
    \node[circle, draw] (x1) at (2, 2) {$x_1$};
    \node[circle, draw] (x4) at (4, 0) {$x_4$};
    \node[circle, draw] (x3) at (2, -2) {$x_3$};

    \draw[->] (x1) to[bend left=20] node[pos=0.6, below] {0.01} (x2);
    \draw[->] (x2) to[bend left=20] node[pos=0.4, below] {0.01} (x1);
    \draw[->] (x1) to[bend left=20] node[pos=0.6, above] {0.01} (x4);
    \draw[->] (x4) to[bend left=20] node[pos=0.4, above] {0.01} (x1);
    \draw[->] (x3) to[bend left=20] node[pos=0.6, below] {1.00} (x4);
    \draw[->] (x4) to[bend left=20] node[pos=0.4, below] {0.02} (x3);
    \draw[->] (x2) to[bend left=20] node[pos=0.6, below] {0.02} (x4);
    \draw[->] (x4) to[bend left=20] node[pos=0.4, below] {0.01} (x2);
    
    \draw[->] (x1) edge[loop above] node {1.01} (x1);
    \draw[->] (x2) edge[loop above] node {1.00} (x2);
    \draw[->] (x3) edge[loop above] node {0.90} (x3);
    \draw[->] (x4) edge[loop above] node {0.05} (x4);
\end{tikzpicture}

    \caption{Weighted network of 4 nodes.}
    \label{Fig6}
    \end{minipage}
    \centering
\end{figure}

A comparative study of three ranking algorithms is presented.
HITS\_PM\_Norm (HITS\_Power Method\_Normalization) is an algorithm that performs the {\textbf{Power Method}} on a primitivity adjusted symmetric matrix: 
$$A_{[i]}(\alpha) = \alpha {L_{[i]}}^TL_{[i]}  + \frac{1-\alpha}{i + 3}{\bf{1}}{{\bf{1}}}^T, 
( i= 0, 1 ),$$ where ${\bf{1}}=(1,...,1)^T$, 
by normalizing $A_{[i]}(\alpha)x$ in $L2$ norm(See section 11.3 of~\cite{ref_prbook}). 
PageRank\_Norm (PageRank\_L2 Normalization) is an algorithm that performs normalization of $L_{[i]}(\alpha)x$ in the $L2$ norm and the power method on a primitive probability matrix $L_{[i]} (\alpha)$.
Our proposed method is a gradient algorithm for a symmetric matrix $A_{[i]}=A_{[i]}(1)={L_{[i]}}^TL_{[i]}$, which does not require L2 normalization and hyperparameter $\alpha$, and it is based on the theorem 5 given in~\cite{ref_yzw}, $A_{[i]}$ and $H_{[i]}=H_{[i]}(1)=L_{[i]}{L_{[i]} }^T$ and their eigenvectors can be obtained by an iterative algorithm. As far as we know, our method is an unprecedented contribution in that it allows the exact step size to be determined from the potential function.

Tables 1 and 2 show the results of the ranking calculations for $A_{[i]}(\alpha)$ or $L_{[i]}(\alpha)$.
In the $L_{[0]}$ and $L_{[1]}$ examples, when the number of nodes is increased from 3 to 4, the three algorithms experience a reversal or separation of rankings. In particular, PageRank\_Norm results in poor ranking separation performance.
For HITS\_PM\_Norm, depending on the setting of the hyperparameter $\alpha$ , no ranking inversion occurs in Table 1, but in Table 2, a ranking inversion occurs.
“b-1” means ‘10 to the power of -1,’ which indicates that the number should be moved one decimal place to the left.

\begin{table}{
\caption{Ranking values for the link matrix $L_0$}
\begin{center}
\scalebox{0.8}[0.9]{ 
\begin{tabular}{ |p{1.0cm}||p{3.0cm}|p{3.0cm}|p{3.0cm}|p{3.0cm}|}
 \hline
 Node    & Our algorithm & HITS\_PM\_Norm \phantom{x} ($\alpha = 0.8$) 
 & HITS\_PM\_Norm \phantom{x} ($\alpha = 0.3$)  
 & PageRank\_Norm \phantom{x} ($\alpha = 0.8$)  \\
 \hline
 $x_1$   & 7.23b-1     & 4.84b-1    &3.56b-1     &3.34b-1      \\
 $x_2$   & 2.77b-1     & 4.11b-1    &3.50b-1     &3.33b-1      \\
 $x_3$   & 4.17b-32    & 1.05b-1    &2.93b-1     &3.33b-1      \\
 \hline
\end{tabular}
}

\label{tab1}
\end{center}


\vspace{30pt} 

\caption{Ranking values for the link matrix $L_1$}
\begin{center}
\scalebox{0.8}[0.9]{ 
\begin{tabular}{ |p{1.0cm}||p{3.0cm}|p{3.0cm}|p{3.0cm}|p{3.0cm}|}
 \hline
 Node & Our algorithm & HITS\_PM\_Norm \phantom{x} ($\alpha = 0.8$) 
 & HITS\_PM\_Norm \phantom{x} ($\alpha = 0.3$)  
 & PageRank\_Norm \phantom{x} ($\alpha = 0.8$)  \\
 \hline
 $x_1$   & 1.17b-4       &1.43b-2     &1.72b-1     &1.82b-1        \\
 $x_2$   & 3.72b-4       &1.55b-2     &1.71b-1     &1.85b-1        \\
 $x_3$   & 4.46b-1       &4.36b-1     &3.10b-1     &6.20b-1       \\
 $x_4$   & 5.53b-1       &5.34b-1     &3.47b-1     &1.26b-2        \\
 \hline
  \end{tabular}
}
\label{tab2}
\end{center}
}
\end{table}

\section{Application to cell production system}
After describing the on-site scene to be understood, the overall flow from detection to analysis of event logs will be explained, and the detection method and the method of analyzing process networks using the event node ranking method will be explained.

\subsection{Cell production site}
\begin{figure}
\centering
\includegraphics[width=0.85\textwidth, keepaspectratio]{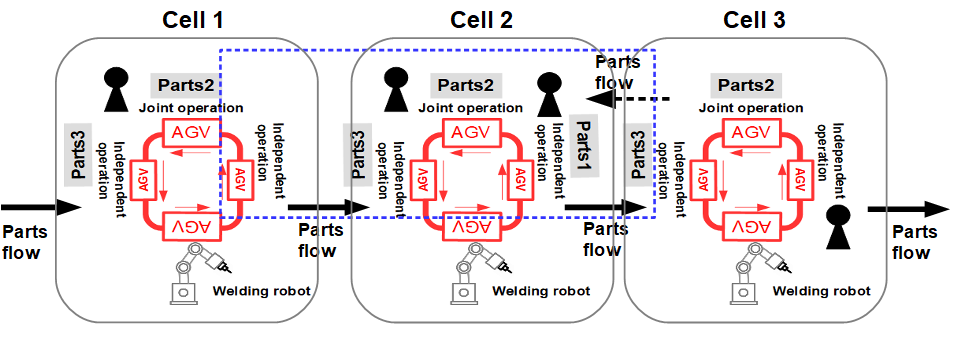}
\caption{Cell production system and target area (blue dotted box).} \label{Fig7}
\end{figure}

\begin{figure}
\centering
\includegraphics[width=1.0\textwidth, keepaspectratio]{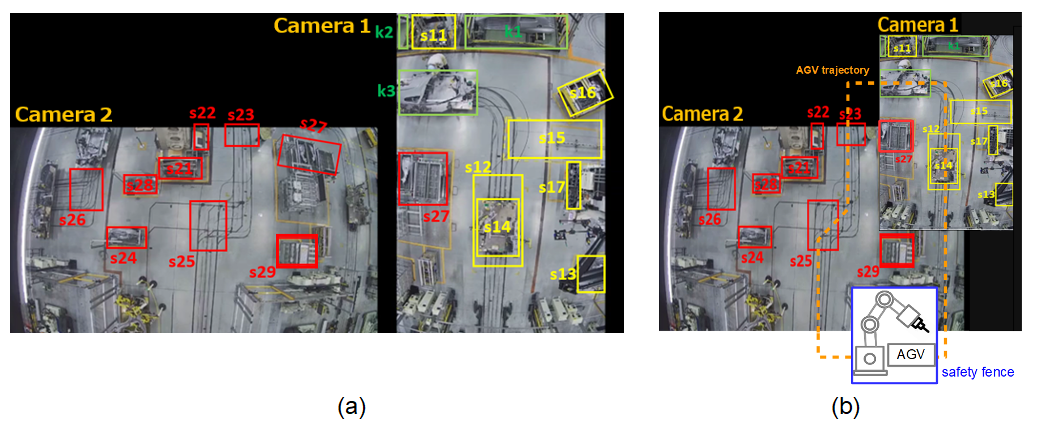}
\caption{(a) Location setting and event detection. (b) Semantic integration of (a).} 
\label{Fig8}
\end{figure}

Figure~\ref{Fig7} illustrates the interconnection among three cells. Cell 2 is situated between Cell 1 and Cell 3 and represents a process with a high workload. Consequently, to share part of the workload of Cell 2, the flow of parts between cells is reversed, enabling Cell 3, the subsequent process, to assist with some of Cell 2's tasks. To closely monitor Cell 2, which could potentially become a bottleneck in the cell production system, two cameras were installed on the ceiling, approximately $7$ meters above the ground, to capture the area within the blue dotted lines in Fig.~\ref{Fig7}. The images enclosed in Fig.~\ref{Fig8}(a) correspond to the locations indicated by the blue dotted lines in Fig.~\ref{Fig7}.

In Cell 2, there are two workers. The work areas are designated as follows: collaborative or individual work areas k1 to k3, shown with green bounding boxes in Figure~\ref{Fig8}; individual work areas s11 to s17, shown with yellow bounding boxes; and individual work areas s21 to s29, shown with red bounding boxes. These assignments are pre-planned. The yellow and green work areas are assigned to Worker A (also referred to as the right-area person), while the red and green work areas are assigned to Worker B (also referred to as the left-area person).

Two large AGVs and two small AGVs circulate within the cell. The presence of both large and small AGVs is to accommodate the preparation and handling of left and right automotive parts, respectively. Assuming that each location corresponds to a specific task, we detect and monitor the operations occurring within the cell.

\subsection{Event detection and visualization}
In order to capture the event via the installed multi-camera, we employ an object detection technique as demonstrated in \cite{ref_Luft}. 
Our fundamental idea for ``lift-up'' object detection to event detection is, to identify the overlap of the region between the detected object (workers or AGVs, for instance) and the predefined area that we are interested in (such as a working bench). 
Indeed, we define ``worker's event occurred'' if the total time of overlap between the bounding box of a stationary work area, such as a parts shelf, and the bounding box of a worker is sufficient (see Figure~\ref{Fig7}).

Rather than typical AI-empowered event detection~\cite{ref_deep}, which directly predicts the predefined event by the whole pixels of the input, thus fragile to ``background'' (such as luminance change, not-intended AGV arrival, etc.), our indirect strategy converts an image into objects of interest and their positions and then identifies their relationships with the environment. 
Therefore, it can more robustly localize the event of interest against the other less important ones.

\begin{figure}
\centering
\includegraphics[width=0.9\textwidth, keepaspectratio]{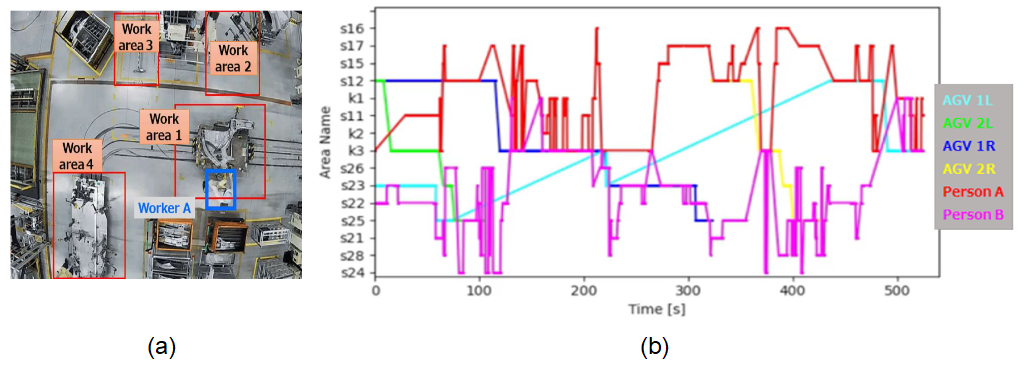}
\caption{(a) Event detection. (b) Gantt chart with $83.1 \%$ precision obtained by our proposal. } \label{Fig9}
\end{figure}

To simplify the event description,  we only capture the starting time of each task for \emph{our} event detection, instead of recording each task's duration time. 
This is because the end of each task indicates the start of the consequent task in almost all cases. 
We named the trajectory of its event start \emph{Gantt Chart} and depict the example hereafter (Fig.~\ref{Fig9} (b)).
Note that, as far as we know, this is the first work that describes how to bridge object detection and event detection into a practical scenario that cooperated with the details of its protocol.

In our experiment, we implemented object detection by YOLOv3~\cite{ref_yolov3}. We manually annotated to train the model via CVAT~\cite{ref_opencv_cvat} and obtained the object detector that predicts the position of six classes (four types of AGV and two types of workers that different jobs are assigned to each) when provided the $416 \times 416$ resized image.
To determine the event, we heuristically chose the following thresholds; {\it {3 second  duration}} to distinguish whether ``event started'' when overlapped; {\it{10 \% overlapping ratio}} of the bounding box to judge whether it was overlapped. Despite its simplicity, we confirmed that our proposal usually achieves the $80\%$ or more in precision metrics~\cite{ref_precision} (Fig.~\ref{Fig9}(b)).
Note that, for this evaluation, our manual annotation of the worker's standard procedures provides the ground truth of the event detection. Therefore, our proposed algorithm is less prone to record meaningless events (such as just walking over the predefined working area) and tends to record the desired event correctly. 

\subsection{Event ranking and analysis}
We show the applicability of our proposal to real factory data (Fig.~\ref{Fig4}) from three process cycles in cell production: two cycles of data that are completed usually ($EL_1$ and $EL_2$), and the other one is from an unusual cycle ($EL_3$). Note that, the “usual or not usual” criterion is simply determined by cycle time $CT (\cdot)$. 
To validate the capability of our proposed analysis purely, we use the event log created by humans through visual inspection of videos. 
The event log provides the following task completion time such that: 
$CT(EL_1) = \text{8 minutes 30 seconds}$,
$CT(EL_2) = \text{8 minutes 24 seconds}$ and  
$CT(EL_3) = \text{10 minutes 38 seconds}$. 

The network nodes generated from each event data are defined by a pair of entities indicated by the following abbreviations and locations defined in Fig.~\ref{Fig8}(a).
\begin{align*}
\rm{LP} & = \text{Person in charge of red/green work areas in left} \\
\rm{RP} & = \text{Person in charge of yellow/green work areas in right}  \\
\rm{P}  & = \text{Persons (two workers)}  \\
\rm{BV} & = \text{Big AGV (two vehicles, undistinguished)}  \\
\rm{SV} & = \text{Small AGV (two vehicles, undistinguished)}.
\end{align*}

We generated the process model as a weighted network system ( We will also call it a process network ) using ProM 6.9 and 'Mine with Directly Follows visual Miner' (MDFM~\cite{ref_prom}). For knowledge on process mining based on both theoretical and practical case studies, refer to \cite{ref_pm}.
Table~\ref{tab1} presents the top 10 activity nodes extracted from the process network generated from the data of $EL_1$, $EL_2$, and $EL_3$. Here "Activities" meams the number of occurrences.

\begin{table}[htbp]
\caption{Nodes Ranking}
\begin{center}
\scalebox{0.9}[1.0]{ 
\begin{tabular}{ |p{1.2cm}||p{2.5cm}|p{2.5cm}|p{2.5cm}| }
\hline
 \multicolumn{4}{|c|}{Ranking and Node Activity(frequency) for Event Log} \\
 \hline
 Ranking & $EL_1$ node & $EL_2$ node & $EL_3$ node \\
 \hline
 1   &  RP\_s14 (8)    & PR\_s11 (9)    &RP\_s11 (11)\\
 2   &  RP\_s11 (8)   & RP\_s14 (8)   &RP\_k3 (8) \\
 3   &  RP\_k3 (7)   & RP\_k3 (6)    &LP\_k3 (7) \\
 4   &  RP\_s15 (6)   & RP\_s15 (6)   &RP\_s15 (7) \\
 5   &  LP\_s21 (6)   & LP\_k3 (5)   &LP\_s21 (6) \\
 6   & RP\_s17 (5)   & LP\_s23 (5)   &LP\_s27 (5) \\
 7   &  LP\_s22 (5)   & LP\_s27 (4)   &LP\_s22 (5) \\
 8   &  LP\_k3 (4)    & P\_k3 (4)   &RP\_s14 (4) \\
 9   &  P\_k3 (4)     & LP\_s21 (4)   &P\_k3 (4) \\
 10  &  LP\_s23 (4)   & LP\_s22 (4)     &LP\_s23 (3) \\
 \hline
\end{tabular}
}
\label{tab1}
\end{center}
\end{table}


When $L_i, (i=1,2,3)$ is the link matrix for the MDFM generated from $EL_i, (i=1,2,3)$, the top 10 ranked nodes in Tables~\ref{tab2} and ~\ref{tab3} are obtained for each $EL_i$.

\begin{table}[htbp]
  \centering
  \begin{minipage}{0.50\textwidth} 
  \centering
  \captionsetup{skip=1pt} 
  \caption{Node Ranking for Authority Matrix $A$}
    \fontsize{8pt}{12pt}\selectfont 
    \scalebox{0.9}[0.9]{ 
    \begin{tabular}{ |p{1.2cm}||p{2.5cm}|p{2.5cm}|p{2.5cm}| }
    \hline
     \multicolumn{4}{|c|}{Ranking and Node Value for Authority Matrix $A$} \\
     \hline
     Ranking & $A_1$ node & $A_2$ node & $A_3$ node \\
     \hline
     1   &  RP\_k3 (7.47b-1)    & PR\_k3 (7.97b-1)    &RP\_s11 (6.53b-1)\\
     2   &  LP\_s27 (3.57b-1)   & RP\_s15 (3.46b-1)   &RP\_k3 (3.96b-1) \\
     3   &  RP\_s11 (3.48b-1)   & LP\_k3 (2.69b-1)    &LP\_k3 (3.80b-1) \\
     4   &  LP\_s23 (2.46b-1)   & LP\_s27 (2.43b-1)   &LP\_s27 (3.41b-1) \\
     5   &  RP\_s15 (2.07b-1)   & RP\_s11 (2.27b-1)   &RP\_s15 (2.51b-1) \\
     6   & BV\_s23(1.37b-1)   & LP\_s23 (1.93b-1)   &LP\_s21 (1.45b-1) \\
     7   &  RP\_s14 (1.32b-1)   & LP\_s21 (7.86b-2)   &LP\_s22 (1.23b-1) \\
     8   &  LP\_k3 (1.26b-1)    & RP\_s14 (6.84b-2)   &SV\_k3 (1.22b-1) \\
     9   &  P\_k3 (1.09b-1)     & SV\_s14 (4.75b-2)   &BV\_s12(1.05b-1) \\
     10  &  RP\_s17 (9.14b-2)   & P\_k1 (4.64b-2)     &P\_k3 (8.40b-2) \\
     \hline
    \end{tabular}
    }
    \label{tab2}
  \end{minipage}
    
 \hspace{20\textwidth} 
 
  \centering
  \begin{minipage}{0.50\textwidth} 
  \centering 
  \captionsetup{skip=1pt} 
  \caption{Node Ranking for Hub Matrix $H$}
    \fontsize{8pt}{12pt}\selectfont 
    \scalebox{0.9}[0.9]{ 
    \begin{tabular}{ |p{1.2cm}||p{2.5cm}|p{2.5cm}|p{2.5cm}| }
    \hline
     \multicolumn{4}{|c|}{Ranking and Node Value for Hub Matrix $H$} \\
     \hline
     Ranking & $H_1$ node & $H_2$ node   & $H_3$ node   \\
     \hline
     1   &  RP\_s11 (8.06b-1)    & RP\_s11 (8.65b-1)    &RP\_k3 (5.29b-1)\\
     2   &  P\_k3 (2.55b-1)   & LP\_k3 (3.16b-1)   &RP\_s11 (4.84b-1) \\
     3   &  LP\_s21 (2.40b-1)   & RP\_s14 (2.56b-1)    &LP\_k3 (4.71b-1) \\
     4   &  RP\_k3 (2.32b-1)   & RP\_k3 (1.59b-1)   &LP\_s27 (3.01b-1) \\
     5   &  LP\_k3 (2.28b-1)   & LP\_s27 (1.43b-1)   &P\_k3 (2.39b-1) \\
     6   &  LP\_s27 (1.94b-1)   & P\_k3 (9.02b-2)   &LP\_s21 (1.65b-1) \\
     7   &  SV\_s14 (1.85b-1)   & LP\_s22 (8.84b-2)   &LP\_s23 (1.34b-1) \\
     8   &  LP\_s22 (1.45b-1)    & LP\_s21 (7.92b-2)   &RP\_s14 (1.25b-1) \\
     9   &  RP\_s15 (9.33b-2)     & RP\_s15 (7.73b-2)   &BV\_k3 (1.23b-1) \\
     10  &  RP\_s17 (8.00b-2)   & SV\_s14 (6.41b-2)     &SV\_k3 (1.15b-1) \\
     \hline
    \end{tabular}
    }
    \label{tab3}
  \end{minipage}
\end{table}

\section{Discussion}
In camera recognition, when the work on the AGV is finished, the transmitter button is pressed, and such quick movements of small hands, which are not actual working time, are excluded from detection in advance and handled. The object detection has been confirmed to be sufficiently useful based on the standard work manual, we were able to detect events in the field that can reproduce the process.

Next, consider it in terms of event node analysis.
The node values in Tables 4 and 5 are squared for each node, and their sum is 1 (normalization). This is a property of our ranking algorithm.
As seen in Table 3, when examining the top 10 nodes from the MDFM activity index, the differences in the nodes that appear are small. This indicates that it is sometimes difficult to determine whether a cycle is usual or unusual based on MDFM activity alone.
The distribution of node values for $EL_3$ in Tables 4 and 5 shows that the variation is greater than the distribution of node values for $EL_1$ and $EL_2$.
This suggests that certain nodes are not in their usual state according to the ranking-based analysis method.

If we look at $EL_1$ and $EL_2$ in terms of cycle time, they are considered to be in their usual cycle time, but,
$EL_3$ is not. This difference can be discovered by ranking analysis of the event nodes without requiring special knowledge.
Looking at matrices $A_1$, $A_2$, and $A_3$ in Table 4, there are seven common nodes in the top 10 for both $A_1$ and $A_2$, and six common nodes in the top 10 for both $A_2$ and $A_3$.
The nodes common to $A_1$ and $A_2$ and not common to $A_3$ are $RP\_s14$ and $LP\_s23$, and both nodes have AGV human work on $s14$ and $s23$.

\begin{itemize}
\item
In A3, it is inferred that AGV work is less accessible.
\end{itemize}

On the other hand, there are nine common nodes in the top decile of the hub matrices of $H_1$ and $H_2$ and seven common nodes in the top decile of the hub matrices of $H_1$ and $H_3$.
The nodes common to $H_1$ and $H_2$ but not to H3 are $LP\_s22$, $RP\_s15$, and $SV\_s14$. Here, $s22$ is the parts shelf, $s15$ is waiting for parts in the back-end process, and $s14$ is AGV work.

From these activities 
\begin{itemize}
\item 
Unlike usual, it can be inferred that several other tasks are being worked on.
\end{itemize}

Thus, by observing the upper layers of the node rank in terms of both the authority matrix and the hub matrix for each cycle,
one can notice which nodes to focus on without requiring special knowledge.
\vspace{0.2cm}

We have examined the first steps of spatial process mining in manufacturing and think it has potential for use in areas other than manufacturing. With the ongoing evolution of generative AI and fundamental 
models, active sensing will expand to image, language, and other modalities, which will be important future research.


\section{Conclusion}
The contributions of this paper can be summarized in the following three points.
\begin{itemize}
\item
A new framework for Spatial Process Mining was proposed.
\end{itemize}

\begin{itemize}
\item
The event logs detected by the camera are visualized in a simple Gantt chart, which facilitates quantitative understanding of the event logs.
\end{itemize}

\begin{itemize}
\item
By considering the ranking of nodes in a complex process network from the viewpoint of authority matrix and hub matrix, we have proposed a method that allows process nodes, which play an important role in complex process networks, to be understood without knowledge of the hyperparameter.
\end{itemize}



\section*{Acknowledgments}
We would like to thank Dr. Hideki Kajima for primitive  discussions of the concept of spatial process mining and also Kazuhiro Shintani, Yuto Mori, Toru Hirai, and Yoshiki Fujiwara for their assistance in data collection and useful discussions. We thank Yuki Kondo for his discussions with us on object detection and for bringing us about research trends in image stitching.


%
%
%

\end{document}